\documentclass[conference, review]{IEEEtran}

\usepackage{graphicx}
\usepackage[english]{babel}
\usepackage{lipsum}
\usepackage{cite}
\usepackage{xcolor}
\usepackage{dblfloatfix}
\usepackage{multirow}
\usepackage[para]{threeparttable}
\usepackage{balance}
\usepackage{import}
\usepackage{hyperref}
\usepackage{enumitem}
\usepackage{color}
\usepackage{booktabs}
\usepackage{bm}
\usepackage[caption=false]{subfig}
\usepackage{adjustbox}
\usepackage{amsmath}
\usepackage{relsize}
\usepackage[ruled,vlined,linesnumbered]{algorithm2e}
\usepackage{setspace}
\usepackage{array}
\usepackage{caption}
\usepackage{subcaption}
\usepackage{fontawesome}

\definecolor{darkgreen}{rgb}{0.4, 0.6, 0.3}
\definecolor{darkgray}{rgb}{0.6, 0.6, 0.6}

\usepackage{listings}
\definecolor{codegreen}{rgb}{0,0.6,0}
\definecolor{codegray}{rgb}{0.5,0.5,0.5}
\definecolor{codepurple}{rgb}{0.58,0,0.82}
\definecolor{backcolour}{rgb}{0.95,0.95,0.95}
\definecolor{black}{rgb}{0,0,0}

\lstdefinestyle{mystyle}{
    backgroundcolor=\color{backcolour},   
    commentstyle=\color{codegreen},
    keywordstyle=\color{magenta},
    numberstyle=\tiny\color{codegray},
    stringstyle=\color{codepurple},
    basicstyle=\ttfamily\footnotesize,
    breakatwhitespace=false,         
    breaklines=true,                 
    captionpos=b,                    
    keepspaces=true,                 
    numbers=left,                    
    numbersep=5pt,                  
    showspaces=false,                
    showstringspaces=false,
    showtabs=false,                  
    tabsize=2
}

\lstdefinestyle{mystyle2}{
    backgroundcolor=\color{backcolour},   
    commentstyle=\color{codegreen},
    keywordstyle=\color{black},
    numberstyle=\tiny\color{codegray},
    stringstyle=\color{codepurple},
    basicstyle=\ttfamily\footnotesize,
    breakatwhitespace=false,         
    breaklines=true,                 
    captionpos=b,                    
    keepspaces=true,                 
    numbers=left,                    
    numbersep=5pt,                  
    showspaces=false,                
    showstringspaces=false,
    showtabs=false,                  
    tabsize=2
}

\SetKw{Continue}{continue}

\ifCLASSINFOpdf
\else
\fi

\AtBeginDocument{\bstctlcite{BSTcontrol}}
\makeatletter
 \def\@IEEEauthorblockconfadjspace{-0.3cm}
\makeatother
\let\svthefootnote\thefootnote
\begin{document}

\title{Alkaid: A Compiler Framework for Ultra-Low-Latency Kernels on Hardware}

\author{Author names removed for blind review\vspace*{-1.2cm}}

\author{
    \IEEEauthorblockN{
        Chang Sun\IEEEauthorrefmark{1}\IEEEauthorrefmark{2},
        Zhiqiang Que\IEEEauthorrefmark{3}\IEEEauthorrefmark{4},
        Dimitrios Danopoulos\IEEEauthorrefmark{5},
        Maurizio Pierini\IEEEauthorrefmark{5},
        Wayne Luk\IEEEauthorrefmark{4},
        Maria Spiropulu\IEEEauthorrefmark{2}
    }

    \IEEEauthorblockA{
        \IEEEauthorrefmark{2}
        California Institute of Technology, USA;
        \IEEEauthorrefmark{3}
        University of Bristol, UK; \\
        \IEEEauthorrefmark{4}
        Imperial College London, UK;
        \IEEEauthorrefmark{5}
        CERN, Switzerland
    }
    \vspace{-1cm}
}

\newcommand{\fmax}{$\mathrm{F}_\mathrm{max}$}

\maketitle

\begin{abstract}

    Ultra-low-latency machine learning and data processing pipelines operating on the sub-microsecond level often contain static dataflow kernels that require fine-grained bitwidth control, arithmetic optimization, and fast hardware performance estimation.
    This works introduce Alkaid, a free and open source domain specific compiler that translates sub-microsecond latency dataflow kernels into platform agnostic Register Transfer Level (RTL) or High-Level Synthesis (HLS)-ready code in seconds.
    Alkaid targets ultra-low-latency ML pipelines, especially those addressed by existing ML-to-hardware flows such as \texttt{hls4ml}, while also supporting the surrounding pre-processing, post-processing, and control logic.
    At its core, Alkaid uses an arithmetic centric intermediate representation, Alkaid Low-level IR (ALIR), to preserve fixed-point semantics, heterogeneous bitwidths, and operation level structure for optimization and code generation. Alkaid further provides comprehensive hardware-aware optimization passes and an interpretable white-box analytical performance model for rapid design space exploration without synthesis in the loop.
    Across representative ML workloads, including MLPs, GNNs, and transformers, Alkaid achieves 20-30\% lower LUT usage than the best prior flows for bit-exact equivalent designs, with up to 40\% lower latency in some cases. Beyond typical neural networks, Alkaid achieves comparable performance to the state-of-the-art method for synthesizing boosted decision trees, while also enabling the implementation and integration of non-ML kernels such as sorting networks and histograms.
\end{abstract}

\IEEEpeerreviewmaketitle

\section{Introduction}

FPGAs are an attractive platform for sub-microsecond, ultra-low-latency computing, especially in domains such as real-time scientific instrumentation~\cite{hls4ml} and edge intelligence.
In these applications, the target computation is often a static dataflow kernel whose graph structure, tensor shapes, and data types are known at compile time. Examples include low-latency machine learning inference kernels and the surrounding data-processing logic used in Level-1 trigger pipelines at the Large Hadron Collider (LHC). Unlike throughput oriented accelerators, these kernels often target sub-microsecond latency, where small changes in bitwidth or arithmetic structure can directly affect both resource usage and timing.

Existing ML-to-hardware frameworks such as \texttt{hls4ml}~\cite{hls4ml} and FINN~\cite{finn} have shown that neural network models can be translated into efficient low-latency hardware.
However, these tools are organized around coarse grained neural network layers with predefined templates. While this makes them effective for supported models, they are less flexible when new ML architectures introduce operators outside the predefined library~\cite{hls4ml}, or when the full application kernel includes custom pre/post-processing operations like tensor arrangement, selection, sorting, or control logic.
To address this gap, we present A Library for Kernel Accelerator Implemented on Device (\textbf{Alkaid}), a free and open source domain specific compiler framework
\let\thefootnote\relax\footnotetext{\textsuperscript{*}Corresponding author. Email: \url{chsun@cern.ch}}
\let\thefootnote\svthefootnote
\footnote{https://github.com/calad0i/alkaid}
for low-latency static dataflow kernels.
Alkaid translates Python/NumPy descriptions and machine-learning frontend models into Alkaid Low-level IR (ALIR), an arithmetic-centric static single-assignment intermediate representation. ALIR preserves fixed-point semantics, heterogeneous bitwidths, and operation-level dependencies throughout tracing, optimization, performance modeling, and code generation.
In addition, Alkaid lowers the optimized ALIR into platform agnostic Register Transfer Level (RTL) or High-Level Synthesis (HLS)-ready code. It also provides an interpretable white-box analytical performance model that estimates area and combinational latency from the optimized IR, enabling rapid design-space exploration without invoking synthesis in the loop. Although our evaluation uses FPGA backend tools, the generated RTL is not tied to a specific FPGA vendor and can be integrated into downstream hardware design flows.

To the best of our knowledge, Alkaid is the best-performing compiler framework for sub-microsecond dataflow, and the first one providing an accurate, white-box analytical performance modeling.
We make the following contributions:
\begin{itemize}[noitemsep,topsep=3pt]
    \item We present Alkaid, an open-source domain-specific compiler framework for ultra-low-latency ML pipelines and their surrounding data-processing logic.
          Alkaid translates python/numpy descriptions and ML frontend models into platform agnostic RTL or HLS-ready code, enabling fast hardware generation for both ML and non-ML kernels.

    \item We introduce ALIR, an arithmetic-centric SSA intermediate representation that preserves fixed-point semantics, heterogeneous bitwidths, and operation-level structure across tracing, optimization, modeling, and code generation. Based on ALIR, Alkaid provides hardware-aware optimization passes and a white-box analytical performance model for rapid area and latency estimation without synthesis in the loop.

    \item We evaluate Alkaid across MLPs, Graph Neural Networks (GNNs), transformers, boosted decision trees (BDT), sorting networks, histograms, and custom python/numpy dataflow kernels.
          For bit-exact neural network designs, Alkaid reduces LUT usage by $20\sim30\%$ and latency by up to 40\% compared with the state-of-the-art (SOTA) \texttt{da4ml} or \texttt{hls4ml}+\texttt{da4ml} flows, while reducing end-to-end compilation and synthesis time from hours or days to under an hour.

\end{itemize}

\section{Background and Related Work}
\label{sec:background}

Existing ML-to-hardware frameworks such as \textit{hls4ml}~\cite{hls4ml}, \textit{chisel4ml}~\cite{c4ml}, \textit{FINN}~\cite{finn} and \textit{fpgaConvNet}~\cite{venieris2016fpgaconvnet} have shown that neural networks can be translated into efficient FPGA implementations. Other frameworks, such as TreeLUT~\cite{treelut}, Conifer~\cite{conifer}, and FwX~\cite{fwx1,fwx2}, have explored the implementation of decision trees and tree ensembles on FPGAs. However, these frameworks are coarse-grained template-based, making it difficult to implement new layers, which requires sufficient hardware knowledge to manually implement the corresponding layer in either RTL or HLS.

More broadly, compiler infrastructures such as Dynamatic~\cite{josipovic2018dynamically}, Calyx~\cite{nigam2021compiler}, CIRCT~\cite{circt2026}, and XLS~\cite{xls2026} reflect the growing
interest in compiler-based procedural hardware generation. In parallel, frameworks such as DnnWeaver~\cite{dnnweaver}, FPGAConvNet~\cite{venieris2016fpgaconvnet}, ScaleHLS~\cite{ye2022scalehls}, HIDA~\cite{ye2024hida}, POM~\cite{zhang2024optimizing}, Allo~\cite{chen2024allo}, and Stream-HLS~\cite{basalama2025stream} have explored automated HLS generation from affine or multi-kernel programs. However, these frameworks are not directly comparable to the proposed framework as their target applications operate at latencies that are orders of magnitude higher than the sub-microsecond regime targeted by Alkaid, where frameworks such as \texttt{hls4ml}~\cite{hls4ml} and \texttt{FINN}~\cite{finn} dominate.

\begin{table}[tb]
    \newcommand{\tworows}[2]{\begin{tabular}[c]{@{}c@{}}#1 \\ #2\end{tabular}}
    \newcommand{\mfend}{\tworows{Multiple}{ML frameworks}}
    \newcommand{\mbend}{\tworows{Multiple}{hardware vendors}}
    \newcommand{\quant}{\tworows{Arbitrary}{Quantization}}
    \newcommand{\yes}{\faCheckSquare}
    \newcommand{\no}{\faSquareO}
    \newcommand{\partially}{\faPlusSquareO}
    \addtolength{\tabcolsep}{-0.3em}
    \caption{Comparison of different tools for the translation of sub-microsecond ML models to FPGA designs.}
    \label{tab:related_work}
    \resizebox*{0.5\textwidth}{!}{
        \begin{tabular}{c|ccccc}
            Framework              & Alkaid & hls4ml                        & da4ml\textsuperscript{1} & FINN                          & Chisel4ML \\
            \midrule
            HAO\textsuperscript{2} & \yes   & \partially\textsuperscript{3} & \yes                     & \no                           & \no       \\
            Extensible Operation   & \yes   & \partially                    & \no                      & \no                           & \no       \\
            Vendor Agnostic        & \yes   & \no                           & \yes                     & \no                           & \yes      \\
            Extensible Frontend    & \yes   & \partially                    & \no                      & \no                           & \no       \\
            MLP/Convolution        & \yes   & \yes                          & \yes                     & \yes                          & \yes      \\
            Transformer            & \yes   & \yes                          & \no                      & \yes                          & \no       \\
            Fast Compilation       & \yes   & \no                           & \yes                     & \no                           & \yes      \\
            Non-NN Operations      & \yes   & \no                           & \no                      & \no                           & \no       \\
            Performance Model      & \yes   & \partially\textsuperscript{4} & \no                      & \partially\textsuperscript{4} & \no       \\
            \midrule
        \end{tabular}
    }

    \yes: Supported; \no: Not supported; \partially: Partially supported
    \vspace{0.5em}

    {
        \footnotesize
        \setstretch{1}
        $^1$ da4ml as a standalone compiler. $^2$ HAO denotes hardware-aware optimization, i.e., optimization using non-trivial explicit hardware-aware transformations that are not delegated to the backend synthesis tool, i.e., beyond pragma tuning and basic algebraic simplifications.
        $^3$ hls4ml can use da4ml as an optimization provider but lacks first-party support.
        $^4$ Only for a subset of operations. \par
    }
    \normalsize
    \vspace{-1.5em}
\end{table}

\section{Design and Architecture}
\label{sec:design}

\begin{figure*}
    \centering
    \includegraphics[width=0.9\textwidth]{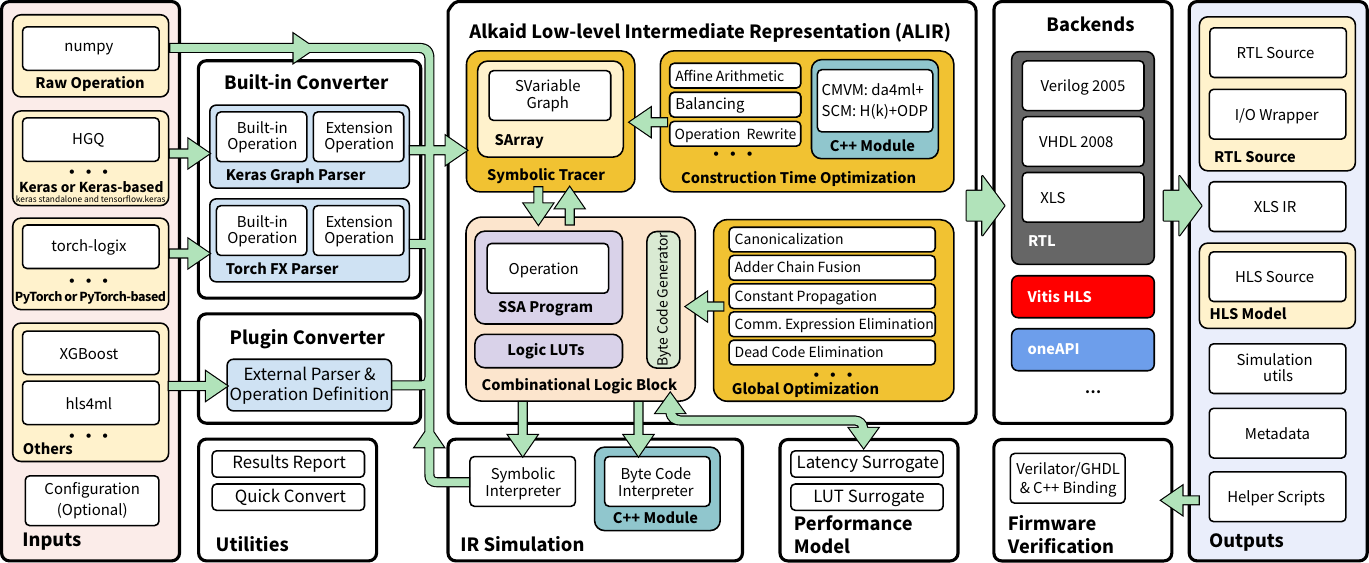}
    \vspace{-0.1cm}
    \caption{
        Architecture and workflow of the Alkaid framework. All components shown, except the frontend frameworks, are developed as part of the framework.
    }
    \vspace{-0.65cm}
    \label{fig:architecture}
\end{figure*}

Alkaid is a domain-specific, kernel-level compiler that translates static dataflow kernels, or computation graphs whose structure and data types are fully determined at compile time, into platform-agnostic RTL or HLS code. To lower the barrier for defining such kernels, Alkaid provides a symbolic tracing frontend based on Python and NumPy: the user ``replays'' the kernel's execution on a symbolic tensor provided by the framework using native arithmetic and NumPy operations, and the framework traces the executed operations to construct the corresponding IR. After the IR is generated, it can be optimized with a set of built-in passes, or the user may define custom transformations for specific applications. RTL or HLS code is then generated from the IR with a code generator, which can be used for direct synthesis or as a template for further manual optimizations. The overall architecture of Alkaid is illustrated in Figure~\ref{fig:architecture}.

\subsection{Intermediate Representation and Optimizations}

\subsubsection{ALIR}

ALIR, the intermediate representation of Alkaid, is an instruction-based, low-level SSA static dataflow IR tailored for DSP-heavy kernels. While many existing IRs are expressive enough to represent our kernels, ALIR is designed to serve not only as a functional representation of computation, but also as a direct optimization and code-generation substrate for low-latency hardware. The key design choice is that ALIR is \emph{arithmetic-centric}: all operations are natively defined on fixed-point numbers with arbitrary bitwidths, and the bit-extensions or concatenations required when operating between values of different bitwidths are implicit in the operation semantics rather than explicit IR instructions. By preserving the arithmetic structure instead of lowering to explicit bit-manipulation, ALIR enables optimizations, such as affine-arithmetic-based interval propagation for bitwidth minimization, that would be significantly more convoluted in a bitvector-centric IR like the XLS IR.

ALIR is partly inspired by Distributed Arithmetic Instruction Set (DAIS) used in da4ml~\cite{da4ml}, but the two serve different purposes. DAIS was mainly introduced for distributed arithmetic, with minimal extensions for supporting the inference of specific neural networks. On the other hand, ALIR is designed as a more general arithmetic-centric IR, covering a wider range of static dataflow kernels beyond DA-style linear operators. This includes comparison, selection, lookup, and bitwise operations, while still preserving explicit SSA dependencies, fixed-point semantics, and heterogeneous bitwidth support. As a result, ALIR retains the low-level arithmetic transparency of DAIS, while providing a compiler abstraction better suited to Alkaid's end-to-end optimizations and direct RTL/HLS generation for a broader range of applications.

Currently, ALIR implements only 13 low-level operations that have been found to sufficiently cover all operations in the dataflow kernels we have encountered in our target applications, including basic arithmetic operations (neural networks), comparison and selection operations (e.g., sorting networks or decision trees), and boolean operations (e.g., logic-gate networks or generic conditional switches). The operations are listed in Table~\ref{tab:perf_model}. For each operation, we map it to a minimal implementation in the target RTL or HLS code, which can be fed into the synthesis tool for downstream processing.

\subsubsection{Optimizations}

While the synthesis backend can also perform most optimizations,
we find that explicitly implementing certain optimizations can significantly reduce the area and latency of the generated design.

Hence, we implemented these optimizations in Alkaid as a part of the compilation process. Individual optimizations may be switched on or off by the user with a simple API, and the user may also define custom optimizations to further optimize the IR for specific applications when needed. In general, for kernels that can reasonably fit into one FPGA ($\sim$1M LUTs), the whole symbolic tracing, IR generation, and optimization process can be completed in milliseconds to under a minute.

The da4ml algorithm included in Alkaid, da4ml+, is a reimplementation of the da4ml algorithm~\cite{da4ml} with optimized memory access patterns for better cache efficiency in C++ with modified tie-breaking rules. This implementation is generally 2.5-3 times faster than the original implementation (e.g., $13.3\rightarrow5.5$ ms for $32\times32\times8$ bits, $84\rightarrow24$ s for $128\times128\times8$ bits random matrices), while achieving the same or better optimization results in adder counts under the same logic depth constraint.

\begin{algorithm}
    \setstretch{0.8}
    \newcommand{\bw}{\mathrm{bw}}
    \newcommand{\bwo}{\mathrm{bw}_{\mathrm{out}}}
    \caption{Lookup operation resource estimator}
    \label{alg:lutdecomp}

    \DontPrintSemicolon
    \SetKwFunction{Count}{Count}
    \KwIn{Logic lookup table $T$ with output bits $\bwo$}
    \KwOut{Estimated LUT count $c$}

    $w \gets \bw(\hat{q})$\;
    \tcp{Bypass slow large table}
    \lIf{$w - X > 6$}
    {\Return $\lceil 0.7 \cdot 2^{w-X} \rceil$}

    $c \gets 0$\;
    \For{$b \gets 0$ \KwTo $\bwo - 1$}{
        $\mathbf{f} \gets \mathrm{bit}\; b \;\text{of}\; T$, reshaped to $\{0,1\}^{w}$\;
        $c \gets c + \Count(\mathbf{f},\, \mathit{false})$\;
    }
    \Return $\lceil c \rceil$\;

    \medskip
    \SetKwProg{Fn}{Function}{:}{}
    \Fn{\Count{$\mathbf{f} \in \{0,1\}^d$, greedy}}{
        \lIf{$d \le X$}{\Return $[\mathbf{f}\text{ is non-const}]$}
        \eIf{\textnormal{greedy}}{
            $i^* \gets \arg\max_i \sum_j [\mathbf{f}_{0,i}[j] = \mathbf{f}_{1,i}[j]]$\;
        }{
            try all $i^* \in [d]$, return minimum\;
        }
        $\mathbf{f}_{k,0} \gets \mathrm{take}(\mathbf{f},\, 0,\, \mathrm{dim}{=}i^*)$\;
        $\mathbf{f}_{k,1} \gets \mathrm{take}(\mathbf{f},\, 1,\, \mathrm{dim}{=}i^*)$\;

        \lIf{$\mathbf{f}_{0,i^*} = \mathbf{f}_{1,i^*}$}{\Return \Count{$\mathbf{f}_{0,i^*}$, \textnormal{true}}}
        \Return \Count{$\mathbf{f}_{0,i^*}$, true} $+$ \Count{$\mathbf{f}_{1,i^*}$, true}\;
    }
    \vspace{-0.1cm}
\end{algorithm}

Due to the low-level nature of the IR, we split optimizations into both the symbolic tracing stage and the traditional IR stage. Optimizations that are local, such as the da4ml optimization for CMVM operations, or optimizations requiring only information from past operations, such as the affine-arithmetic-based interval estimation and some constant folding/propagation, are performed during the symbolic tracing stage as the IR is being constructed. Optimizations requiring a global view of the IR, such as expression canonicalization, common sub-expression elimination, dead/null operation elimination, and further constant folding after these optimizations, are performed iteratively after the IR is constructed. This way, we can maximize the optimization opportunities while minimizing the number of iterations required for global optimizations.

\begin{table*}[htbp]
    \newcommand{\buf}[1]{\mathrm{buf}[\mathrm{id}_{#1}]}
    \newcommand{\bw}{\mathrm{bw}}
    \newcommand{\bwi}{\mathrm{bw}_{\mathrm{in}}}
    \newcommand{\bwo}{\mathrm{bw}_{\mathrm{out}}}
    \newcommand{\bwii}{\mathrm{bw}_{\mathrm{in},0}}
    \newcommand{\bwij}{\mathrm{bw}_{\mathrm{in},1}}
    \newcommand{\bwred}{\mathrm{reduc}\left(\bwii,\bwij\right)}
    \newcommand{\tworow}{\multirow{2}{*}}
    \setlength\extrarowheight{0.3pt}
    \caption{ALIR operations with their corresponding surrogate resource and latency models. Notation is defined below.}
    \label{tab:perf_model}
    \begin{adjustbox}{width=\textwidth}
        \begin{tabular}{ccllll}
            \hline
            \textbf{Opcode} & \textbf{Name}             & \textbf{Description}                 & \textbf{Cost (LUTs)}              & \textbf{Latency}           & \textbf{Note}               \\
            \hline
            $-2$            & \texttt{NEG}              & Negate value                         & $0$                               & $0$                        & absorbed                    \\
            $-1$            & \texttt{READ}             & Load from external buffer            & $0$                               & $0$                        & no-op                       \\
            \hline
            \tworow{$0/1$}  & \tworow{\texttt{ADD/SUB}} & \tworow{$\buf{0}\pm\buf{1}\ll s$}    & $\max(\bwo-1,1)$                  & $(\bwo-1)\cdot a+b$        & has overlap                 \\
                            &                           &                                      & $0$                               & $0$                        & no overlap                  \\
            \hline
            $2$             & {\texttt{RELU}}           & {$\max(0, x)$}                       & $e \cdot \bwi$                    & $0$                        &                             \\
            \hline
            $3$             & \texttt{QUANT}            & Quantize (bit-take/extend)           & $0$                               & $0$                        & no-op                       \\
            \hline
            \tworow{$4$}    & \tworow{\texttt{CADD}}    & \tworow{$\buf{0} + c \ll s$}         & $d \cdot\max(\bwi-1,0)$           & $0$                        & consumer $\in\{0,1,6\}$     \\
                            &                           &                                      & $0$                               & $0$                        & otherwise (absorbed)        \\
            \hline
            $5$             & \texttt{CONST}            & Define constant value                & $0$                               & $0$                        & no-op                       \\
            $6$             & \texttt{MUX}              & 2:1 mux on MSB of third operand      & $\bw(q)\cdot(0.5 - d\cdot\alpha)$ & $0.5 - g\cdot\alpha$       & Rough approx                \\
            $7$             & \texttt{MUL}              & $\buf{0} \times \buf{1}$             & $\bwred$                          & $\bwred\cdot a+b$          & Rough approx                \\
            \hline
            $8$             & \texttt{LOOKUP}           & Table lookup                         & See Algorithm~\ref{alg:lutdecomp} & $\max(\bwi-6, 1)\cdot 0.5$ &                             \\
            \hline
            $9$             & \texttt{UNARY}            & Bitwise NOT / reduce-OR / reduce-AND & $0$                               & $0$                        & absorbed                    \\
            \hline
            \tworow{$10$}   & \tworow{\texttt{BINARY}}  & \tworow{Bitwise AND/OR/XOR}          & $0.2\cdot\bw_\mathrm{overlap}$    & $h$                        & has overlap                 \\
                            &                           &                                      & $0$                               & $0$                        & no overlap or const operand \\
            \hline
        \end{tabular}
    \end{adjustbox}

    \medskip
    \small
    \noindent\textbf{Notation.}
    All costs are in LUT6 equivalents; latencies are in combinational logic levels, independent of the pipelining configuration. $a\approx 0.025, b\approx 1.1, c\approx 0.36, d\approx 0.3, e\approx 0.667, f\approx 0.7, g\approx 0.4, h\approx 0.15$ are empirically derived constants based on micro-benchmarks. The bitwidth of the input is $\bwi$, or $\bwii, \bwij$ for binary operations, and the output bitwidth is $\bwo$. $\bw_\mathrm{overlap}$ is the overlapping bitwidth of the two operands. $\alpha \in [0,1]$ is the \emph{constant-descendant score} for the \texttt{MUX} operation, which describes the portion of the source terms of the mux-tree (i.e., casted muxing operations) that are constants. \texttt{NEG}, \texttt{WRAP}, and \texttt{BINARY} are considered transparent in the descendant score tracing. ``Consumer $\in \{0,1,6\}$'' refers to if the const-add operation's output is used by any one of the \texttt{ADD}, \texttt{SUB}, or \texttt{MUX} operations. When the cost of an operation is $0$ with certain conditions, we consider the operation to be absorbed into the neighboring operations without adding extra resource overhead, or the operation is a pure signal rearrangement without arithmetic operation. $\mathrm{reduc}(x,y)=x\cdot y - \max(x,y)$ is the estimator of the number of adders used to implement a multiplication with bitwidths $x$ and $y$ implemented as conditional shift-adders.
    \vspace{-0.6cm}
\end{table*}

\subsection{Performance Model}

After these optimizations, the IR closely reflects the post-technology-mapping physical design, enabling a white-box analytical model to estimate area and latency without synthesis.
For each IR operation, we combine first-principles analysis, such as LUT pin counting, with empirical fitting from micro-benchmarks to estimate its LUT cost and combinational delay.
Unlike the machine learning based surrogate models used in hls4ml~\cite{wa-hls4ml,rule4ml}, this analytical approach is directly derived from the IR structure and the underlying hardware implementation, making it interpretable and straightforward to adapt to different target platforms or synthesis tools by adjusting the model parameters. The evaluation of the performance model typically completes in under a second per kernel, which we consider negligible within a typical development cycle.

Table~\ref{tab:perf_model} shows the equations used to estimate LUT usage and latency for each operation.
For area, we model only LUT consumption, since DSP usage is negligible after distributed-arithmetic optimization, while FF usage depends mainly on pipelining rather than the implemented function.
The LUT-decomposition function for the \texttt{LOOKUP} operation is described in Algorithm~\ref{alg:lutdecomp}, which is implemented as a recursive function that decomposes the logic lookup table into two subtables until the length is smaller than $2^6$ for LUT6 on UltraScale+ devices.
Since LUT6\_2 fracturing is rarely used for \texttt{LOOKUP}, fractional LUT usage is not included in the model.

\subsection{Symbolic Tracing Frontend}

The symbolic tracing frontend is Alkaid’s primary frontend, allowing users to define dataflow kernels using native Python and numpy operations on framework-provided symbolic tensors. It is designed to be bit-exact: operations on symbolic tensors match those implemented in hardware. Intermediate bit-growth is tracked using affine arithmetic where possible, with interval arithmetic as a fallback. Therefore, no implicit quantization is applied; users must explicitly use properly quantized operations and weights, otherwise the generated firmware may become unsynthesizable due to excessive intermediate value sizes. In addition, this frontend also underpins higher-level frontends for machine-learning frameworks. For Keras and PyTorch models, Alkaid traces execution on symbolic tensors for most native operations, while framework-specific plugins define how special layers or operators map to NumPy-based symbolic operations. For other frameworks, Alkaid provides a generic interface for developers to define the full symbolic execution flow.

The plugin system decouples compiler infrastructure from frontend development, allowing each frontend to evolve independently and add new operators within its own codebase. It also allows subtle framework-specific differences, such as padding semantics or quantized lookup behavior, to be represented through different symbolic mappings without forcing compromises in the backend.

\subsubsection{Layer/Operation Plugin}

Currently, Alkaid natively supports the Keras and PyTorch frontends for most of the elementary operations. However, since the symbolic tracing frontend does not perform implicit quantization, the models will need to be explicitly quantized before being fed into the framework. In practice, we expect these blocks to be used in conjunction with third-party libraries providing the mapping of their quantization schemes to the symbolic tracing frontend. Currently, plugins for the HGQ~\cite{hgq}, PQuantML~\cite{pquant}, and torchlogix~\cite{torchlogix} are implemented with the Keras and Torch frontends.

\begin{figure*}
    \centering
    \includegraphics[width=0.99\textwidth]{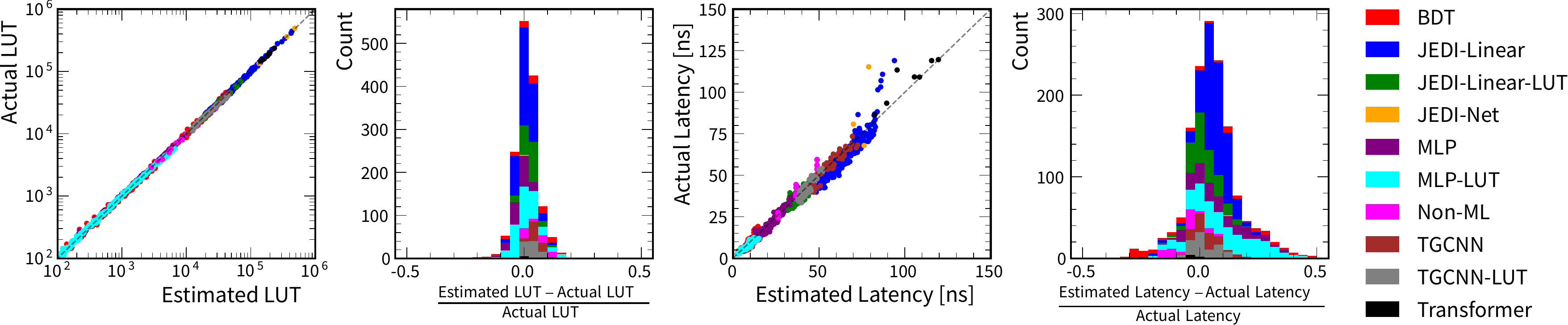}
    \vspace{-0.2cm}
    \caption{
        From left to right: (1) Estimated LUT usage vs. actual LUT usage after place-and-route for various designs. (2) Relative error of the estimated LUT usage vs. actual LUT usage after place-and-route. (3) Estimated latency vs actual latency after place-and-route. Here, latency refers to the real implementation latency after place-and-route with timing requirements met, rather than the number of clock cycles. Larger, more complex JEDI-net designs exhibit higher prediction errors due to increased routing congestion. (4) Relative error of the estimated latency vs actual latency after place-and-route.
    }
    \vspace{-0.6cm}
    \label{fig:perf_model}
\end{figure*}

\subsubsection{Framework-wise Plugin}

When the frontend model is provided by another framework, such as XGBoost~\cite{xgb} or hls4ml~\cite{hls4ml}, developers may define a plugin describing how the model should be executed on the symbolic tensors. This plugin is generally more flexible than the prior operation/layer plugin, while also requiring more effort to implement. For instance, the XGBoost plugin would allow the user to configure how the leaf values are mapped/quantized to fixed-point numbers, and if the tree is to be implemented as a lookup-table or muxes. The hls4ml plugin allows the conversion of an optimized hls4ml model targeting the Vitis HLS backend into ALIR, which can then be further lowered into RTL or HLS code with the Alkaid code generator and remains bit-exact to the original hls4ml model.

\subsection{Code Generation Backend}

Currently, Alkaid provides two code generation backends: a platform-agnostic RTL generator, whose output can be used for direct synthesis or as a template for further manual integration, and an HLS generator targeting Vitis HLS and oneAPI DPC++. Alkaid also provides an IR-to-IR translator layer that emits XLS IR from the optimized ALIR via the C binding to libxls, enabling integration with XLS-based flows. The code generation backend is designed to be extensible to other target platforms by implementing the corresponding code generator and adjusting the performance model accordingly.

\subsection{hls4ml Integration}

While Alkaid is mainly designed as a standalone tool, an alternative to \texttt{hls4ml}, for users who are already familiar with hls4ml, or have existing models defined in hls4ml, Alkaid provides seamless integration with hls4ml to allow users to benefit from Alkaid's broader operator support and faster development cycle while retaining the familiar hls4ml interface. In particular, for the hls4ml/Keras frontend, when an unsupported layer is encountered during hls4ml's parsing, the hls4ml plugin for Alkaid will be triggered to convert the unsupported layer into ALIR. The ALIR will then be converted into HLS code with Alkaid's code generator and inlined into the generated HLS code from hls4ml, which can then be fed into the synthesis tool for downstream processing.

\section{Evaluation}

Unless otherwise stated, all Alkaid results are obtained using Vivado 2025.1 targeting the UltraScale+ VU13P FPGA (\texttt{xcvu13p-flga2577-2-e}), based on out-of-context post-routing reports. Following prior work, we report latency as the number of cycles at the maximum achievable clock frequency (\fmax). All designs are registered at both input and output ports to cover all combinational paths, and the cycle latency is measured as the number of registered combinational stages.

In all evaluated neural network designs, Alkaid generates bit-exact RTL designs compared to the original designs, while reducing the LUT usage by 20-30\% and latency by up to 40\% compared to da4ml or hls4ml+da4ml implementations. Comparing to hls4ml standalone, it achieves 50-70\% LUT reduction and $>99\%$ DSP reduction without the \texttt{da4ml} algorithm optimization. Compared to TreeLUT for decision tree based models, Alkaid achieves comparable efficiency in LUT usage and latency.

\begin{table*}
    \vspace{-0.2cm}
    \caption{
        Performance and resource usage on the JSC HLF dataset. Alkaid reduces LUT utilization by 20\%-30\% compared with da4ml. Models with identical accuracy are converted from the same trained model and have bit-exact firmware implementations.
    }
    \newcommand{\nrow}[2]{\multirow{#1}{*}{#2}}
    \begin{adjustbox}{width=0.73\textwidth,center=\textwidth}
        \begin{threeparttable}
            \setlength\extrarowheight{-0.8pt}
            \begin{tabular}{llc|cccccc}
                \midrule
                Model          & Implementation                 & Acc. $\uparrow$  & Latency & LUT    & DSP & FF     & \fmax [MHz] & II \\
                \midrule
                \nrow{10}{MLP} & \textbf{Alkaid}                & \nrow{2}{76.9\%} & 27.8 ns & 7,341  & 0   & 8,532  & 540.        & 1  \\
                               & FPGA'26~\cite{hgq} HGQ (da4ml) &                  & 36.3 ns & 10,182 & 0   & 10,480 & 552.        & 1  \\
                               & \textbf{Alkaid}                & \nrow{2}{76.6\%} & 21.2 ns & 4,423  & 0   & 4,727  & 565.        & 1  \\
                               & FPGA'26~\cite{hgq} HGQ (da4ml) &                  & 26.2 ns & 5,991  & 0   & 5,890  & 611.        & 1  \\
                               & \textbf{Alkaid}                & \nrow{2}{75.6\%} & 15.8 ns & 1,744  & 0   & 1,740  & 569.        & 1  \\
                               & FPGA'26~\cite{hgq} HGQ (da4ml) &                  & 18.6 ns & 2,298  & 0   & 2,217  & 645.        & 1  \\
                               & \textbf{Alkaid}                & \nrow{2}{74.9\%} & 12.6 ns & 1,001  & 0   & 909    & 556.        & 1  \\
                               & FPGA'26~\cite{hgq} HGQ (da4ml) &                  & 14.5 ns & 1,390  & 0   & 1,316  & 692.        & 1  \\
                               & \textbf{Alkaid}                & \nrow{2}{73.2\%} & 7.1 ns  & 271    & 0   & 288    & 708.        & 1  \\
                               & FPGA'26~\cite{hgq} HGQ (da4ml) &                  & 9.1 ns  & 366    & 0   & 363    & 663.        & 1  \\
                \midrule
            \end{tabular}
        \end{threeparttable}
    \end{adjustbox}
    \label{tab:jsc-hlf}
    \vspace{-0.3cm}
\end{table*}

\subsection{Performance Modeling}

To evaluate the accuracy of the performance model, we compare the estimation from the performance model with the post-routing results from Vivado 2025.1 for a large set of machine learning models, including the multi-layer perceptron (MLP), static graph neural networks (GNN), JEDI-linear~\cite{jedi-linear}, TGCNN~\cite{tgc}, and transformers. Some non-ML kernels, such as the parallel sorting networks, are also included in the evaluation since they are commonly used in the LHC trigger pipelines. A total of 2,410 designs are evaluated for the performance model, with a wide range of LUT usage from half a thousand to over half a million.

We show the correlation plots and the distribution of relative estimation error for both LUT usage and latency in Figure~\ref{fig:perf_model}. We find that our performance model can achieve an estimation error of $<10\%$ for LUT usage and $<20\%$ for latency compared to the post-routing results from Vivado 2025.1 in general. In both cases, we find other factors, such as the target clock frequency, routing congestion, or even the version of the synthesis tool may lead to a larger variation in the actual area and latency of the generated design, which is not and cannot be captured by the performance model. This appears to be a fundamental limitation shared by any pre-routing analytical model, as even the vendor tools' own timing estimates degrade significantly under heavy routing congestion. Hence, we consider the estimation error of the performance model to be sufficient for users to quickly iterate on their design and explore the design space before running the synthesis step.

\begin{table*}
    \caption{
        Performance and resource usage on the JSC PLF dataset. Alkaid reduces LUT utilization by 20\%-30\% compared with hls4ml using the da4ml plugin, and achieves much lower LUT usage than standard hls4ml.
    }
    \newcommand{\nrow}[2]{\multirow{#1}{*}{#2}}
    \begin{adjustbox}{width=0.92\textwidth,center=\textwidth}
        \begin{threeparttable}
            \setlength\extrarowheight{-0.8pt}
            \begin{tabular}{llcc|lcccccc}
                \midrule
                Model                 & Implementation                              & P            & Acc. $\uparrow$  & Latency  & LUT     & DSP   & FF   & \fmax [MHz] & II & Logic Synth       \\
                \midrule
                \nrow{4}{JEDInet}     & \textbf{Alkaid}                             & \nrow{2}{16} & \nrow{2}{68.9\%} & 108.1 ns & 413 k   & 5     & 172k & 167.        & 1  & \textbf{46.7 min} \\
                                      & [TRETS'26]~\cite{metamlpro} MetaML (hls4ml) &              &                  & -        & 896 k   & 5,013 & 516k & -           & 2  & $>$ 29 h          \\
                                      & \textbf{Alkaid}                             & \nrow{2}{32} & \nrow{2}{74.6\%} & 90.1 ns  & 314 k   & 5     & 152k & 178.        & 1  & \textbf{36.1 min} \\
                                      & [TRETS'26]~\cite{metamlpro} MetaML (hls4ml) &              &                  & -        & 1,017 k & 845   & 761k & -           & 2  & $>$ 41 h          \\
                \midrule
                \nrow{4}{Linformer}   & \textbf{Alkaid}                             & \nrow{2}{16} & \nrow{2}{72.8\%} & 102.3 ns & 185 k   & 0     & 91 k & 186.        & 1  & \textbf{28.9 min} \\
                                      & [ML4PS'25]~\cite{fast-xfm} hls4ml+da4ml     &              &                  & 104.6 ns & 246 k   & 0     & 66 k & 182.        & 1  & $>$ 9 h           \\
                                      & \textbf{Alkaid}                             & \nrow{2}{32} & \nrow{2}{78.4\%} & 86.8 ns  & 198 k   & 0     & 95 k & 196.        & 1  & \textbf{29.1 min} \\
                                      & [ML4PS'25]~\cite{fast-xfm} hls4ml+da4ml     &              &                  & 140.0 ns & 267 k   & 0     & 71 k & 150.        & 1  & $>$ 24 h          \\
                \midrule
                \nrow{4}{Transformer} & \textbf{Alkaid}                             & \nrow{2}{16} & \nrow{2}{72.3\%} & 97.6 ns  & 229 k   & 0     & 99 k & 184.        & 1  & \textbf{37.3 min} \\
                                      & [ML4PS'25]~\cite{fast-xfm} hls4ml+da4ml     &              &                  & 98.4 ns  & 279 k   & 0     & 80 k & 193.        & 1  & $>$ 12 h          \\
                                      & \textbf{Alkaid}                             & \nrow{2}{32} & \nrow{2}{77.0\%} & 79.5 ns  & 141 k   & 0     & 69 k & 189.        & 1  & \textbf{26.2 min} \\
                                      & [ML4PS'25]~\cite{fast-xfm} hls4ml+da4ml     &              &                  & 82.9 ns  & 180 k   & 0     & 55 k & 193.        & 1  & $>$ 30 h          \\
                \midrule
            \end{tabular}
        \end{threeparttable}
    \end{adjustbox}
    \label{tab:jsc-plf}
    \vspace{-0.55cm}
\end{table*}

\subsection{Performance and Resource Usage}

We evaluate the performance of Alkaid on a set of representative machine learning models, including the multi-layer perceptron (MLP), static graph neural networks (GNN), and transformer/linformer models for the jet substructure classification task. For this task, we compare the performance and resource usage of the designs generated by Alkaid with the exact same models used in the respective original works, which are generated with da4ml~\cite{da4ml}, hls4ml~\cite{hls4ml}, or a combination of the two. In all cases, the models converted with Alkaid are bit-exact to the original RTL/HLS code generated by the respective tools, hence there is strictly no difference in the accuracy of the generated designs. For BDTs, we use FQTree~\cite{fqtree}, which leverages Alkaid for hardware generation, and compare agaist TreeLUT~\cite{treelut} on the same task and the MNIST~\cite{mnist} dataset.

For the jet substructure classification (JSC) task, the objective is to classify the jets into 5 categories. We consider two variants of this task, one with only the high-level features (HLF) and the other with the particle-level features (PLF). For the HLF variant, we consider models trained and evaluated on the OpenML variant of the task~\cite{openml-jet}, which contains 16 high-level features derived from the kinematic properties of the jet constituents. For the particle-level variant, we adopt models trained on the hls4ml jet tagging dataset~\cite{hls4ml-dataset}, which contains the kinematic variables of individual particles. For this study, we consider the models with a maximum of 16 or 32 particles with both the GNN and transformer/linformer architectures.

For the decision tree models, we consider the models trained on the same JSC task with the same dataset as well as the MNIST~\cite{mnist} dataset. Since the exact models with the specified accuracy are not provided in the original TreeLUT paper, we fit the decision trees ourselves targeting the same accuracy. The decision trees are then converted into RTL with Alkaid for the evaluation. This results in slightly different models between the two implementations, but we consider the accuracy to be close enough for a meaningful comparison of the performance and resource usage of the generated designs. For this comparison, it should be noted that the latencies reported for TreeLUT~\cite{treelut} are adjusted due to different evaluation protocols. TreeLUT uses unregistered inputs and outputs with 0.05-0.1 ns external delay, while the latency cycle number is counted by the number of internal registers between the input and output ports. On the other hand, we count the number of cycles in Alkaid with the number of combinational blocks between the registered input and output ports, which has +1 cycle compared to the TreeLUT counting method with the same pipelining. To normalize the latency results, we adjust the latency for TreeLUT to be $\text{reported\_latency} + \text{clock\_period} - 0.2\text{ ns}$, where the clock period is calculated as the inverse of the reported \fmax. The 0.2 ns is the higher estimate derived from the 0.05-0.1 ns delays added by the authors for the unregistered input and output ports, which yields the most favorable adjusted latency for TreeLUT. Since TreeLUT omits the input and output registers, the FF counts cannot be directly compared between the two implementations but are included for completeness.

\begin{table*}[htbp]
    \centering
    \newcommand{\nrow}[2]{\multirow{#1}{*}{#2}}

    \caption{Performance and resource consumption of BDT models on the JSC HLF dataset and the MNIST dataset. Alkaid achieves similar resource usage and latency compared to TreeLUT~\cite{treelut} at the same or slightly higher accuracy.}
    \label{tab:bdt}

    \begin{adjustbox}{width=0.70\textwidth,center=\textwidth}
        \begin{threeparttable}
            \setlength\extrarowheight{-0.8pt}
            \begin{tabular}{ll|clccccc}
                \midrule
                Task              & Implementation                 & Acc. $\uparrow$ & Latency [cycle] & LUT   & DSP & FF    & \fmax & II \\
                \midrule
                \nrow{4}{JSC HLF} & \textbf{Alkaid}                & 75.6\%          & 2 (3.8 ns)      & 2,181 & 0   & 566   & 521.  & 1  \\
                                  & FPGA'25~\cite{treelut} TreeLUT & 75.6\%          & *3 (3.9 ns)     & 2,234 & 0   & 347   & 735.  & 1  \\
                                  & \textbf{Alkaid}                & 74.7\%          & 1 (2.0 ns)      & 682   & 0   & 146   & 512.  & 1  \\
                                  & FPGA'25~\cite{treelut} TreeLUT & 74.6\%          & *2 (2.1 ns)     & 796   & 0   & 74    & 887.  & 1  \\
                \midrule
                \nrow{4}{MNIST}   & \textbf{Alkaid}                & 96.6\%          & 2 (3.5 ns)      & 4,429 & 0   & 1,390 & 565   & 1  \\
                                  & FPGA'25~\cite{treelut} TreeLUT & 96.6\%          & *3 (3.6 ns)     & 4,478 & 0   & 597   & 791   & 1  \\
                                  & \textbf{Alkaid}                & 96.0\%          & 2 (3.4 ns)      & 3,306 & 0   & 1,247 & 581   & 1  \\
                                  & FPGA'25~\cite{treelut} TreeLUT & 95.6\%          & *3 (3.2 ns)     & 3,499 & 0   & 759   & 874   & 1  \\
                \midrule
            \end{tabular}
            \vspace{-0.6cm}
            \footnotesize
            \normalsize
        \end{threeparttable}
    \end{adjustbox}
    \vspace{-0.3cm}
\end{table*}

The results for the JSC HLF and JSC PLF tasks are summarized in Tables~\ref{tab:jsc-hlf} and \ref{tab:jsc-plf}, respectively. The results for the decision tree models are summarized in Table~\ref{tab:bdt}. The MLP, transformer, and Linformer models are converted via the HGQ plugin, while the GNN models are converted via the hls4ml plugin for bit-exactness. The BDT models are converted with the XGBoost plugin with explicit quantization support.

For neural network models, we find that Alkaid can generally achieve 20-30\% lower LUT usage compared to the designs generated by \texttt{da4ml} or \texttt{hls4ml} with the da4ml plugin. On some designs, the latency could also be reduced by up to 40\% compared to the original designs, while maintaining similar latency on others. We attribute the LUT improvement over da4ml to the affine arithmetic based bit-growth tracking, which allows Alkaid to use smaller bit-widths for the intermediate values without causing overflow, while da4ml relies on a more conservative interval arithmetic that leads to larger bit-widths in general. The latency improvement is likely from both the reduced arithmetic complexity and better latency performance modeling, which allows for better path length balancing and more efficient pipelining. However, with the backend compiler retiming, we do observe minor fluctuations in latency. Compared to hls4ml without using da4ml for CMVM kernel optimization, Alkaid can achieve significantly lower LUT usage while mostly eliminating DSP usage. For the BDT models, Alkaid achieves equivalent or slightly better accuracy compared to TreeLUT~\cite{treelut} with similar LUT usage and latency after adjusting for the different evaluation protocols.

It should be noted that, for the JEDInet models, we disabled the da4ml optimization for the CMVM kernels to bit-exactly match the original design's accumulator overflow behavior in hls4ml, and hence, the DSP-to-LUT conversion optimization used only the H(k)+ODP SCM optimization, which is expected to be less effective than the full CMVM optimization. However, the Alkaid designs still outperform the original hls4ml designs with $50$-$70\%$ lower LUT usage and $>99\%$ lower DSP usage in general, while simultaneously achieving II=1 compared to hls4ml's II=2, which is strictly higher throughput with lower resource utilization. Due to the high LUT usage, we observe high routing congestion (level 4-5) for the JEDInet designs, which leads to a significant reduction of $\mathrm{F}_\mathrm{max}$ and hence higher latency than the performance model's estimation. Since some designs span over one SLR and the cross-SLR routing delay is not modeled in the performance model, the pipelining is likely to be less efficient than the estimation as well, which also contributes to the lowered \fmax. Since the original hls4ml designs are not routed, we cannot compare the actual latency of the two implementations.

Another advantage of Alkaid over hls4ml, or other HLS-based flows is the significantly faster development cycle enabled by the quick model-to-RTL conversion and accurate performance estimation. For the JEDInet models and the transformer/linformer models, the designs generated by hls4ml, with or without the da4ml plugin, require hours to days for a single HLS synthesis. In particular, the JEDInet models require days for HLS and logic synthesis, while the transformer/linformer models could also need more than one day. In contrast, the model-to-RTL conversion with Alkaid takes less than a minute for all of the models shown, and the logic synthesis for the each take no more than 50 minutes. This allows users to iterate on their designs much more efficiently and explore the design space with more freedom.

\vspace{-0.05cm}

\begin{lstlisting}[emph={with},emphstyle={\bfseries}, language=Python, caption={An example Python function with various NumPy operations that can be automatically traced and converted into RTL code with Alkaid. Non-linear functions (sin fused with multiplication here) are implemented as logic look-ups with precision evaluations lazily evaluated at the next quantization operation. The generated RTL code is bit-exact to the original Python function.}, label=code:legitimate_kernel]
def operation(x): # x[16,4]
    a = np.maximum(x, 0)
    b = x[:, 1::2].T
    c = np.round(np.sin(b) * np.pi)
    d = np.repeat(c, 2, axis=0) * 3 + 4
    e = np.max(np.stack([d, -d*2], axis=0), axis=0)
    f = c @ e.T
    g = np.einsum('ij,kj->ik', c, d)
    h = np.sum(g, axis=0)
    idx = np.argsort(h)[:2]
    j = h[idx].ravel()
    return j[None] + f[..., None] + a[-2::-4, :2]

inp = FVArray.new((8,4)).quantize(0,8,0).as_new()
out = operation(inp)
alir = trace(inp, out)

\end{lstlisting}
\vspace{-0.5cm}

\subsection{Non-Machine Learning Kernels}

Beyond machine learning models, Alkaid can also be used to convert other types of computation kernels into RTL code as long as they can be expressed in reasonably well-structured Python/NumPy code. For instance, one can create a parallel sorting network with the \texttt{np.sort} or \texttt{np.argsort} functions, or a histogram kernel with the \texttt{np.histogram} function. We show the latency and resource usage of the generated designs for these kernels in Table~\ref{tab:non-ml}. We also show an example of how the different NumPy functions can be composed in a pythonic way to create a more complex dataflow kernel in Listing~\ref{code:legitimate_kernel}. In this particular example, the kernel is generated with the inputs being unsigned 8-bit integers, and the generated design will be bit-exact to the description in the Python function, with the intermediate values tracked automatically.
To demonstrate support for a backend beyond AMD FPGAs, these non-ML kernels are synthesized with Quartus 25.1 targeting the Altera Agilex 7 FPGA.

The sorting and histogram kernels are commonly used in the pre-processing stage of many machine learning models at the LHC, such as for jet constituent ordering and feature extraction or creation for the downstream models. With Alkaid, one can create a unified dataflow graph containing both the machine learning model and the pre-/post-processing steps, and convert the entire graph into an RTL project with a single workflow.

\begin{table}[htbp]
    \caption{Latency and resource usage for the sorting networks, histogram, and custom python function generated kernels. The target FPGA is the Altera Agilex 7 \texttt{AGFB014R24A2E2V}.}
    \newcommand{\nrow}[2]{\multirow{#1}{*}{#2}}
    \begin{adjustbox}{width=0.487\textwidth,center=0.487\textwidth}
        \begin{threeparttable}
            \setlength\extrarowheight{-0.8pt}
            \begin{tabular}{ll|rcccccc}
                \midrule
                Kernel                   & Configuration                        & Latency & ALUT   & DSP & FF     & \fmax [MHz] \\
                \midrule
                \nrow{3}{Batcher Sorter} & $128\times 8$ bit                    & 42.1 ns & 41,226 & 0   & 32,845 & 309.        \\
                                         & $96\times 8$ bit                     & 40.2 ns & 29,868 & 0   & 27,145 & 323.        \\
                                         & $64\times 8$ bit                     & 25.3 ns & 15,197 & 0   & 17,302 & 394.        \\
                \midrule
                \nrow{3}{Bitonic Sorter} & $128\times 8$ bit                    & 42.4 ns & 49,432 & 0   & 45,812 & 307.        \\
                                         & $96\times 8$ bit                     & 42.1 ns & 35,170 & 0   & 19,391 & 309.        \\
                                         & $64\times 8$ bit                     & 32.7 ns & 18,525 & 0   & 12,894 & 306.        \\
                \midrule
                \nrow{4}{Histogram}      & $128\rightarrow 8$ bin               & 5.6 ns  & 2,297  & 0   & 739    & 356.        \\
                                         & $256\rightarrow 8$ bin               & 8.4 ns  & 5,006  & 0   & 1,945  & 358.        \\
                                         & $128\rightarrow 32$ bin              & 5.7 ns  & 11,324 & 0   & 8,122  & 348.        \\
                                         & $256\rightarrow 32$ bin              & 9.3 ns  & 24,379 & 0   & 6,651  & 321.        \\
                \midrule
                Custom Dataflow          & Listing~\ref{code:legitimate_kernel} & 6.3 ns  & 3,744  & 0   & 3443   & 326.        \\
                \midrule
            \end{tabular}
        \end{threeparttable}
    \end{adjustbox}
    \label{tab:non-ml}
    \vspace{-0.4cm}
\end{table}

\section{Conclusion}

We present Alkaid, a domain-specific compiler that translates low latency kernels into platform-agnostic RTL or HLS code. Alkaid combines a Python/NumPy symbolic tracing frontend, an arithmetic-centric IR (ALIR), and a white-box analytical performance model for rapid design-space exploration. Across MLPs and transformers, Alkaid reduces LUT usage by $20$-$30$\% and latency by up to 40\% compared with the SOTA da4ml or hls4ml+da4ml flows for bit-exact designs. For GNNs, Alkaid reduces LUT usage by $50$-$70\%$ and DSP usage by $>99\%$ compared with hls4ml standalone, while reducing end-to-end compilation and synthesis time from hours or days to under an hour.
Alkaid also support for non-ML static dataflow kernels such sorting networks and histogram kernels for LHC trigger pre-processing.
As a design choice, Alkaid generates stateless pipelined designs similar to XLS functions, and we consider scheduling and orchestration to be an orthogonal step at the higher level. We aim to enable processing element reuse and throughput-area trade-offs for larger models and support more complex dataflow graphs in the future.

\vspace{0.2cm}
\noindent \textbf{Acknowledgements}
Partial support from the United States DoE (grant numbers DE-SC0011925, DE-FOA-0002705), NSF (grant numbers PHY240298, PHY2117997), United Kingdom EPSRC (grant numbers UKRI256, EP/V028251/1, EP/N031768/1, EP/S030069/1, and EP/X036006/1), AMD, and Eric \& Wendy Schmidt Fund via CERN NGT (agreement number SIF-2023-004), is acknowledged.
\vspace{-0.2cm}

\bibliographystyle{IEEEtran}

\bibliography{biblio}

\balance

\end{document}